# The Simulation of High Pressure Nucleation Experiments in Diffusion Cloud Chamber

**Sergey P. Fisenko**

*A.V. Luikov Heat &Mass Transfer Institute, National Academy of Sciences, Minsk, Belarus*

E mail: fsp@hmti.ac.by

For high- pressure nucleation experiments in upward diffusion cloud chamber, there is the great deviation of predictions of classical nucleation theory from experimental results; the discrepancy is more than 10 orders of magnitude of nucleation rate. Experimental data for 1-propanol vapor are under investigation in this paper. It was shown that mathematical model of a single droplet growth and motion semi- quantitatively explained all experimentally discovered regularities. For explanations low nucleation rate versus high supersaturation, the coalescence mechanism in gaseous phase has been proposed. As result of coalescence the vast majority of newly formed clusters evaporate and restore vapor density and temperature profile in DCC. The observed picture with low nucleation rate is result of diffusion interaction between small clusters and droplets in nucleation zone for high-pressure nucleation experiments.

## Introduction

Diffusion cloud chamber (DCC), shown in Fig.1, is the device widely used in nucleation research, approximately 50% of all nucleation rate and critical supersaturation data in literature were obtained using that device. During high- pressure nucleation experiments [1-3], it was discovered great deviation of the experimental results from predictions of classical nucleation theory [4]. At experiments described at [1-3], the pressure of background (carrier) gas was about (1 ÷ 40) bars. Hydrogen and helium were used as background gases. High- pressure nucleation experiments with nitrogen and argon as background gases are described in paper [5], qualitatively the similar results have been obtained. In particularly, at [1, 2] the dependence of critical supersaturation versus background gas pressure and temperature was established. The most surprising result is that for very high values of supersaturation low nucleation rate have been observed. The relationship of calculated nucleation rate to experimental one can be order $10^{10}$ and higher.

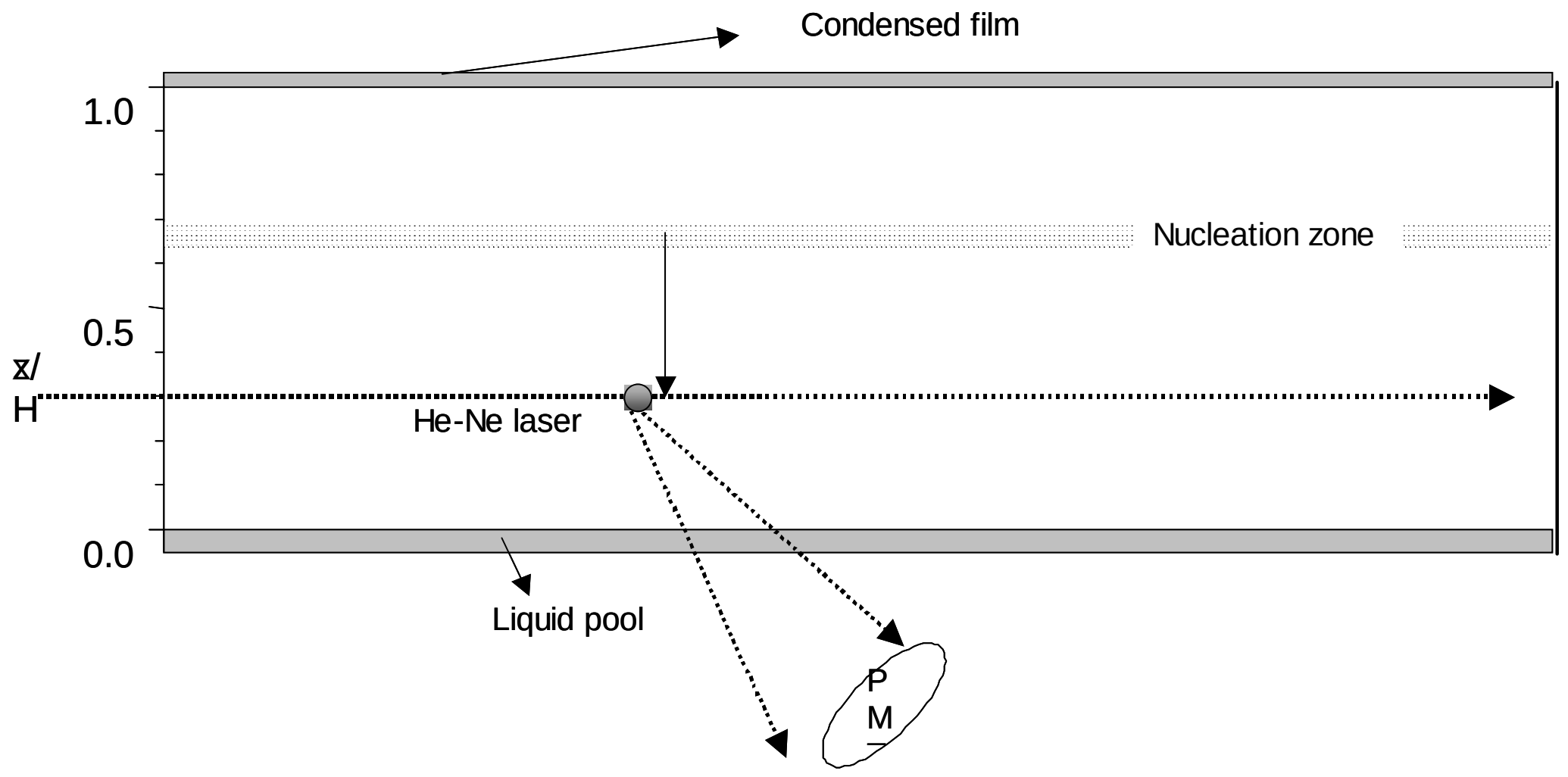

Figure 1 The sketch of diffusion cloud chamber
*H* is the chamber height.

For relatively small pressure of background gas, pressure effect was investigated in [6 and references therein]. Pressure of background gas P affects on the value of diffusion coefficient of vapor D (D~1/P), and correspondingly, affects on the rate of droplets growth at the regime of continuous medium. For optical methods of droplets registration, the intensity of scattering optical signal depends on square of the droplet radius (in the approximation of geometrical optics). Another words, it was suggested that the increasing of pressure does not change the nucleation rate, but change the intensity of scattering optical signal. This effect takes place for high- pressure nucleation experiments also and has to be taken into account for correct interpretation of experimental results.

On the basis of methods of nonequilibrium statistical thermodynamics [7], high- pressure nucleation kinetics was investigated in [8]. It was shown, that if critical cluster is smaller than the mean free path of vapor molecule, there is significant deviation from results of classical nucleation theory [4]. To mention that experimental setup for testing nucleation kinetics developed in [8] has not been made yet.

The aim of this paper is to explain the main puzzle of the high- pressure nucleation experimental results in DCC related with low nucleation rate, at least, semi- quantitatively.

## Experimental results

The basic experimental results have been published in [1, 2, 5]. The homogeneous nucleation rate is dramatically reduced as the total pressure increases. The effect of increasing total pressure on nucleation is different depending whether helium, hydrogen, nitrogen, or argon is used as background gas. It was observed also, the dependence of the critical supersaturation on total pressure (as manifested by the slope of the critical supersaturation vs. total pressure data at constant temperature) is temperature dependent. It was found that the magnitude of this effect becomes greater as the molecular weight of condensable increases. Similarly, for all investigated alcohols this effect becomes greater as the formula weigh of background gas increases, and the slope of the dependence of the critical supersaturation on total pressure increases as the temperature in nucleation zone decreases.

For registration of new phase droplets, the visual observations have been used. It should be remarked that the height in DCC corresponding to the calculated, according classical nucleation theory [4] maximum in the nucleation rate, corresponded to reasonably well with the observed height of the nucleation plate in high pressure DCC. It is worth to note that an observer sees the droplet through the thick quartz wall of the chamber, it means that their radius should be not smaller than $25\times10^{-6}$m [9].

## Results of simulation of droplet growth and motion

For reckoning of profiles of vapor density and temperature, the Galerkin’s solution of diffusion and heat diffusivity equation is used ([10] and Appendix 2). Advantage of the Galerkin’s type solution is it's simple analytical form and quite high accuracy. The dependence of diffusion coefficient on temperature profile is taken into account. The temperature profile affects on the heat diffusivity coefficient also, so iteration procedure is used for solution the system of two partial differential equations with variable coefficients. For example, the vapor density profile *n*(z) with good accuracy can be written as following

$$n(z) = n(0) + \frac{\left(n(H) - n(0)\right)z}{H} + An_1 \sin\left(\frac{\pi z}{H}\right)$$

where n(0) and n(H) are saturated vapor densities at temperatures of bottom and top plates of DCC, respectively. The coefficient $An_1$ is defined as the result of the solution of algebraic equation in Galerkin's method [10] (see also Appendix 2). To mention that $An_1 < 0$. For simplicity, we neglect the contributions of more high modes here as the Defour effect. Figure 2 shows the contour plot of the dimensionless free energy of cluster formation $\Delta\Phi(g,z)/kT(z)$ of propanol for experimental conditions from [1]. For capillary approximation, the dimensionless free energy of cluster formation from g molecules at position with coordinate z is written as following

$$\frac{\Delta\Phi(g,z)}{kT(z)} = -g\ln(S(z)) + \frac{4\pi R(g)^2 \sigma(T(z))}{kT(z)} \quad . \qquad (1)$$

where *R*(g) is the radius of cluster from g molecule, $\sigma(T(z))$ is the surface tension coefficient depending on the temperature, *k* is Boltzmann's constant.

As seen in Fig. 2, there is the saddle point of the surface of the dimensionless free energy of cluster formation $\Delta\Phi(g,z)/kT(z)$. The saddle point of the dimensionless free energy of cluster formation $\Delta\Phi(g,z)/kT(z)$ can found as the result of the solution the following system of equations

$$\partial_z\left(\frac{\Delta\Phi(g,z)}{kT(z)}\right) = \partial_g\left(\frac{\Delta\Phi(g,z)}{kT(z)}\right) = 0 \, .$$

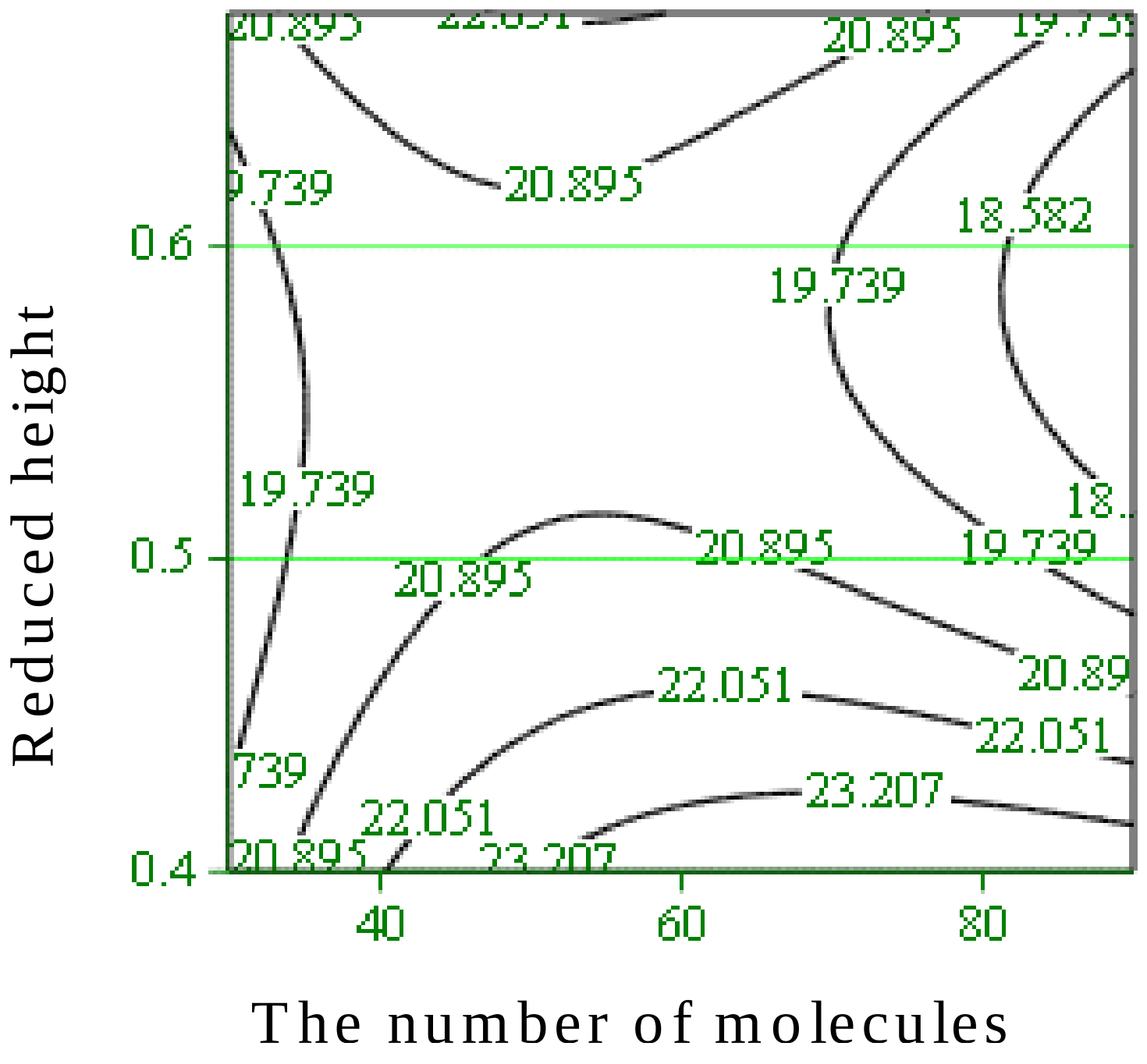

**Figure 2. The surface plot of the dimensionless free energy of cluster formation ΔΦ(g,z)/kT(z) in the cloud chamber**

Abscissa axis is the number molecule into a cluster, ordinate axis is the reduced height of DCC. Experimental conditions: Tb=415.9K, Tu=335K, P=3.5 106 Pa. The saddle point parameters: gc=51.1, zc/H=0.57, ΔΦ(gc,zc)/kT(zc)=20.56 ; Tb is the temperature of the bottom plate of chamber, Tu is the temperature of the upper plate.

We define, follow general ideas of statistical mechanics, the nucleation zone as the spatial domain where the free energy of cluster formation differs from the saddle point value on the one kT(zc). Here T(zc) is the temperature at the saddle point of the chamber. For spatially non- uniform environment of DCC, the width of nucleation zone $h_n$ is about (0.1÷0.15) *H*.

In high pressure nucleation experiments the background gas number density $n_g$ is much large the number density of vapor, therefore to introduce the Knudsen number $Kn_v$ for vapor molecules interacting with new phase cluster (droplet) with radius *R* as following

$$Kn_v = \frac{\lambda_v}{R} = \frac{1}{R n_g a_v \sqrt{1+M/m}} \tag{2}$$

where $\lambda_v$ is the mean free path and $a_v$ is the transport cross –section of vapor molecule with the background gas, M and m are molecular weights of the vapor and gas, respectively. The transition to growth processes at the regime of continuous medium begins for $Kn_v$ <1.

For illustration to mention, if total helium pressure is equal to $25\times10^5$ Pa and temperature 400K, the mean free path of propanol molecule $\lambda_v$ is about $1.2\times 10^{-8}$m. As critical nucleus radius is about $10^{-9}$m, it is mean that for such condition nucleation has to run in free molecular regime ($Kn_v>>1$). Therefore from theoretical point of view, experimental results should not deviate significantly from classical theory of nucleation kinetics in gases [4]. To emphasize, that the classical theory of nucleation kinetics considers mass exchange of new phase cluster with vapor only at the free molecular regime. It is important to remarked that following growth of droplets to optically observed size will be mostly at regime of the continuous medium, when $Kn_v<1$. The basic mechanism of the mass transfer is vapor diffusion. At regime of continuous medium, there are gradients of vapor density near every droplet. For relatively high droplet concentration, some diffusion interactions can be expected between droplets.

Some results of the numerical simulation of droplet growth and motion in diffusion cloud chamber are presented in Fig. 3-4 and at Appendix 1. The chamber parameters and thermodynamic conditions of experiments are taken from [1]. The mathematical model of a droplet growth and motion, used for numerical calculations, is presented in [6]. For simulation, the initial position of droplet coincides with to the position of the saddle point of dimensionless free energy of cluster formation $\Delta\Phi(g,z)/kT(z)$. The initial droplet radius is calculated from thermodynamic conditions at the saddle point. During growth the droplet position versus time is shown in Fig. 3. As seen in Fig. 3, it takes about 0.1 second in order growing droplet begins to fall from nucleation zone. It's not shown in Fig. 3, but our calculation shows, that even droplets with radius about $10^{-6}$m move up under the thermophoretic force. The temperature gradient is very high at high- pressure nucleation experiments [1-3, 5]. It is well known the influence of thermophoretic force is inversely proportional to pressure P, directly proportional to the gradient of temperature and to the square of radius of the droplet. The velocity of droplets is very low at this stage; therefore we can consider them as stagnant ones. The gravity becomes the main governing force only for larger clusters as it proportional to the cube of a droplet radius.

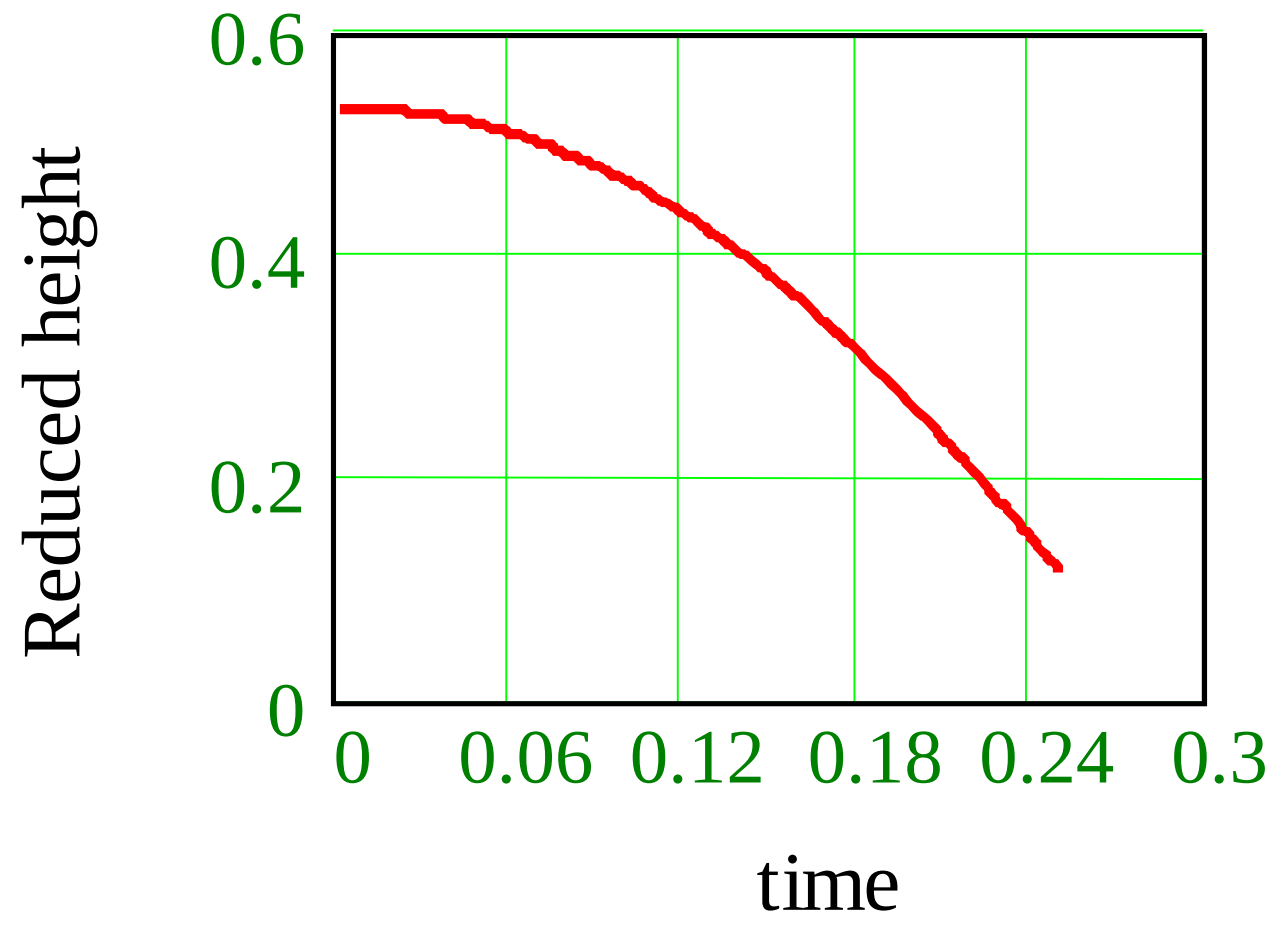

**Figure 3. Dimensionless droplet position versus time of the falling for pressure 39 bars**

In Fig. 4 the change of droplet radius versus time is displayed. To note, the droplet radius is about $3\times 10^{-5}$m at the moment when droplet leaving the nucleation zone. At this point, our numerical results are very close to visual observations in [1, 2], as the position of saddle point practically coincides with the position of maximum calculated nucleation rate. For case when total pressure is about 40 bars, diffusion regime of droplet growth (R(t)~$\sqrt{t}$ ) is obvious from the calculation results, which are presented in Fig. 4.

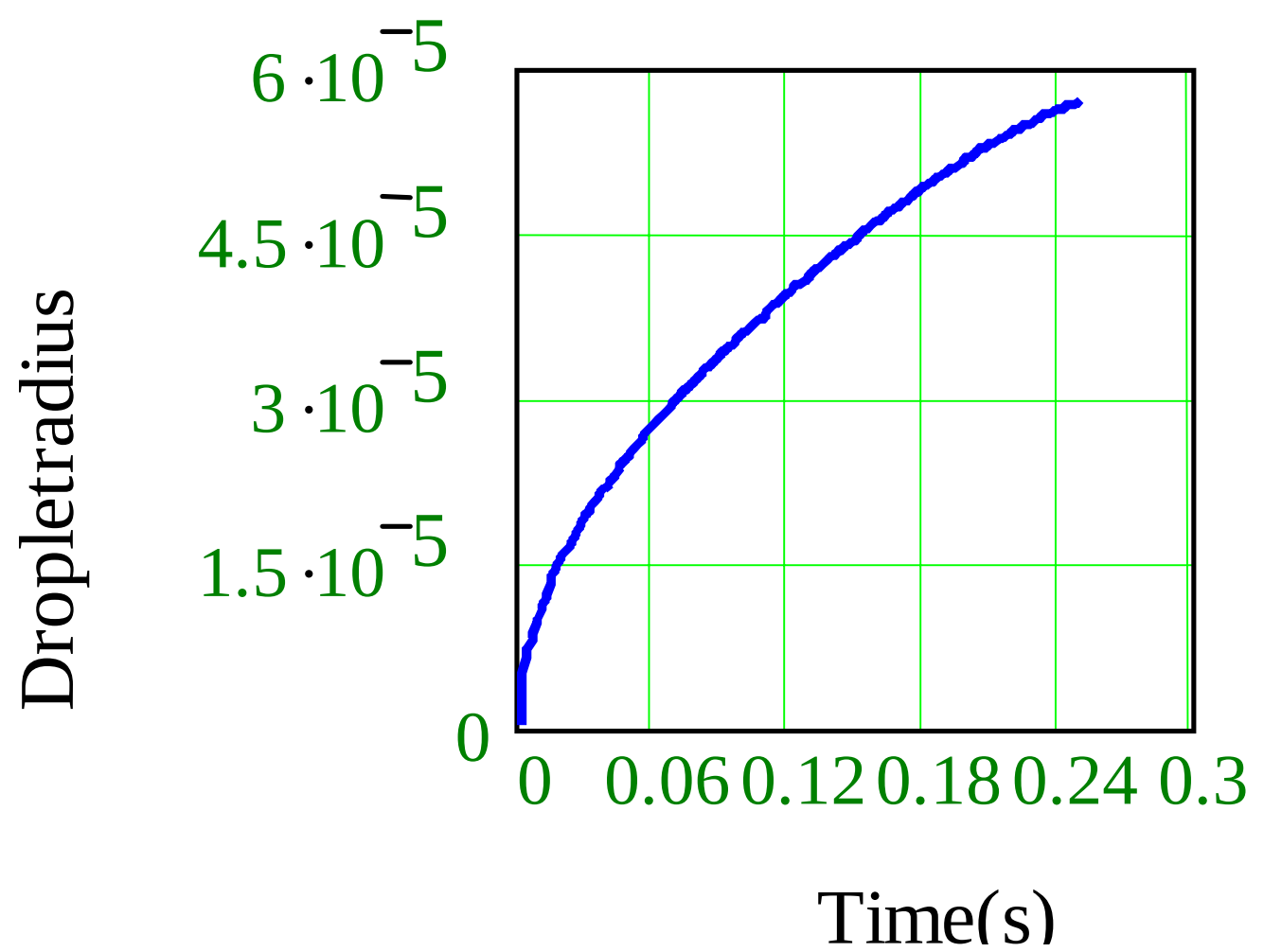

**Figure 4 . Droplet radius R versus time t of falling for pressure 39 bars**

**Other parameters of DCC are presented in [1] for this set of data.**

To use the qualitative estimation [11] of droplet radius R, obtained in [6]

$$R^4 \sim \frac{\eta(T(zc))(S_m - 1)n_e(T(zc))}{\rho^2 g_a P}, \qquad (3)$$

where η(T(zc)) and n(T(zc)) are, correspondingly, the viscosity of background gas and the density of saturated vapor at temperature which is corresponded to temperature for saddle point, $S_m$, maximum of supersaturation, ρ is the density of liquid under investigation, $g_a$ is the gravity acceleration. The intensity of light scattered by droplets, I, is proportional to $R^2$. As shown in Appendix 1 for all experiments described in [1], the droplets radius at the middle of chamber height is practically the same (32÷36 micron). For the observer, the optical signal was practically the same value also. Taking into account the constant value of a optical signal, we have the following expression

$$\sqrt{\frac{\eta(T(zc))(S_m - 1)n_e(T(zc))}{P}} \cong \text{const} \qquad (4)$$

For constant temperature at nucleation zone it follows from (4), that in order to receive the same value of optical signal, the increasing of supersaturation (more exactly, ($S_m$-1)) has to be directly proportional to the increasing of pressure of carrier gas P.

To consider the problem of the correlations between of molecular parameters of a vapor and background gas and the slope of experimental line supersaturation versus pressure, which was discovered in [1, 2, 5]. For high- pressure nucleation experiments, approximate expression for the diffusion coefficient of vapor in background gas can be written as [12]

$$D \sim \frac{kT}{P\sigma_v}\sqrt{\frac{kT}{m_v}},$$

and approximate expression for the viscosity η of background gas can be written as following

$$\eta \sim \frac{\sqrt{m_b kT}}{\sigma},$$

where σ is the transport cross-section of molecules of background gas, $\sigma_v$ is transport cross –section of vapor molecules with background gas molecule, $m_b$ and $m_v$ are, correspondingly, mass of background gas molecule and vapor one.

The generalization of the expression (4) can be received from equations of droplet growth and motion [6]. After some transformations with expressions for transfer coefficients above, the slope of the line, defining the dependence of supersaturation on total pressure in DCC, can be presented as

$$S_m - 1 \sim \text{const} \times P \times \frac{1}{\eta} \frac{\sigma_v \sqrt{m_v}}{(kT)^{3/2} n_s(T)} \ .$$

The use explicit dependence of a viscosity on molecular parameters [12], we have useful expression

$$S_m - 1 \sim \text{const} \times P \times \frac{\sigma_v \sigma}{(kT)^2 n_s(T)} \sqrt{\frac{m_v}{m_b}} \qquad (5)$$

It is evident from expression (5), as mass and the transport cross- section of vapor molecule increases, the slope increases also. Also it has to be remarked, the transport cross section is roughly proportional to mass of molecule. This conclusion confirms the experimental observations [5].

From theoretical point of view, the situation with the correlation between slope of critical supersaturation and the molecular weight of background gas is slightly more complicated. It is clear from (5), that there is strong dependence on transport cross-section of background molecule, and weak inverse dependence on molecular weight. Using the expression (5) and the strong correlation between mass of molecule and its transport cross-section, it is possible to state that the slope increases as molecular weight increases. Qualitatively, the same conclusion was made on the basis of the results of the vast experimental investigations [2, 5].

From expression (5), the temperature dependence of this slope is obvious. To remind only, that of the number density of saturated vapor is exponentially increasing with the temperature increases.

Emphasize that the results of simulation of the droplet growth and motion have been obtained at so-called "one-droplet approximation". This approximation can be used only if the average distance between droplets d is much larger than their radius.

## Diffusion effects near droplet for $Kn_v$ <1

In high-pressure nucleation experiments [1-3, 5], the observable nucleation rate (~1droplet/(s cm$^3$)) was very low contrary to relatively high supersaturation of vapors and, therefore, to classical nucleation kinetics theory. We have here the main puzzle of high-pressure nucleation experiments in diffusion cloud chamber. How it can be solved?

Accordingly the classical nucleation theory for high supersaturation in nucleation zone of chamber, huge number of new phase clusters arises. For condition of high- pressure nucleation experiments in DCC, our calculations show that clusters grow and reach size of the regime of diffusion growth ($Kn_v$<1) in very short time (about ~$10^{-7}$s for radius $10^{-8}$m). For clusters with radius~$10^{-8}$m, diffusion process affects on vapor and non-uniform vapor profile arises near droplet. For propanol cluster with radius R~$10^{-8}$m in He-vapor mixture, characteristic time of vapor diffusion is about $10^{-9}$s. Therefore for estimations, we can use formulas obtained for steady-state regime.

But this regime of droplets growth has not stability against some disturbances. Assume that, occasionally, one of droplets begins to grow faster. The collision of any two clusters can be the origin of faster growth or owing to any statistical fluctuation. In the neighborhood of this relatively large droplet, due to coalescence phenomena, smaller clusters begin to evaporate. As the result of this evaporation process near one growing droplet, the number of droplets decreases. On other side after initial vapor depletion, related with nucleation and fast droplets growth, the steady- state profiles of vapor density and temperature are restored. It was shown above in chapter 1 of this paper, these initial profiles of vapor density and temperature permit to explain many experimental data remarkably well. Emphasize that thermophoretic force and gravity keep clusters practically motionless in nucleation zone during quite long time. The radius of evaporating droplet is about $10^{-8}$ m, when the Kelvin's correction affects on evaporation rate. We use coalescence mechanism for explanation of the main puzzle of high-pressure nucleation experiments.

The coalescence mechanism was applied for investigation only for liquid and solid solutions [13], but high -pressure nucleation experiments create similar conditions at gaseous phase! One important conclusion from coalescence phenomena should be mentioned also. The metastable mixture forgets the initial nucleation rate and its relationship with supersaturation in nucleation zone. The observable variable is only an effective nucleation rate.

We make some quantitative estimation below. Using the classic steady-state solution of the diffusion problem to spherical droplet, it is easily find the formula for supersaturation at distance r from a droplet center

$$S(r) = S_{\infty}\left(1 - \frac{R}{r}\right) + \frac{R}{r}$$

where R is the droplet radius, $S_{\infty}$ is the supersaturation far from the droplet surface. To note, that the supersaturation on the droplet surface is equal to 1. For this case the characteristic time of diffusion is about ($R^2$/D)~ 0.01s for spherical domain with radius R ~ $10^{-4}$m. To mention that this time is much smaller than the residence time of growing large droplet in nucleation zone (see previous chapter). So the application of steady-state formulas is justified.

It follow from the formula above, the effect of local decreasing supersaturation is significant for distance smaller than 10 R. If occasionally new phase clusters arise inside this spatial domain (10R), it will be disappeared due to coalescence mechanism.

The residence time of growing cluster at nucleation zone depends on total pressure of background gas and its nature, as it affects as the viscosity of gaseous mixture as on diffusion coefficient of vapor in background gas.

We make the estimation of diffusion effects near growing droplet on nucleation rate. To consider the cylinder with radius 10R (calculated radius for observable droplet $R\approx3\times10^{-5}$ m, see Fig.4) and height is equal to size of nucleation zone, let the cylinder axis passes with the droplet center. To remind that such droplet is in the nucleation zone about 0.1s. Then, for example, for nucleation rate $10^{22}$ droplet/m$^3$s, about $3\times10^{12}$ droplets will be created during 0.1 second in this cylinder. Again, we assume that classical nucleation kinetics theory works reasonably well (with accuracy of several orders of magnitude). But due to diffusion interaction between growing droplets and large droplet, and, finally, coalescence, we will observe only single stable growing droplet in this cylinder. Therefore, this proposed mechanism decreases the huge discrepancy between formal theoretical result and experimental one. At the example considered above, it decreases roughly on the twenty orders of magnitude of nucleation rate. For larger supersaturation, the larger discrepancy between observed nucleation rate and calculated one could be overcome using the proposed mechanism of diffusion interaction between droplets (coalescence).

In principle, there are four independent parameters in the problem under investigation. The first one is d, the average distance between growing droplets. The second one is $\lambda_v$, the mean free path of vapor molecule, and the third parameter is R, the radius of optically detectable droplet. The fourth parameter is $d_n$, the distance between critical clusters in nucleation zone of DCC.

In order to use the mathematical model of single droplet growth and motion in DCC [6], the following inequality has to be valid

$$d/R>>1 \qquad (6).$$

According to the experimental observations [1-3, 5], the condition (6) is always applicable. For high-pressure nucleation experiments in DCC, the Knudsen number $Kn_v$, defined by expression (2),

$Kn_v<1$,

for propanol droplet with radiuses large than ~$10^{-8}$m. At standard case for nucleation studies in DCC $d_n \cong d$. To mention that quantitatively parameter $d_n$ can be estimated as

$$d_n \sim (\tau I)^{-1/3}, \qquad (7)$$

where I is the steady state nucleation rate for given supersaturation, $\tau$ is characteristic nucleation time about $10^{-6}$s.

In order to the physical picture, described above, was correct, we have to demand the correctness of the following inequality

$$d_n \leq 10\lambda_v \qquad (8).$$

Combining (7) and (8), we receive the estimation of nucleation rate, which consistent with our understanding of physical nature of the experiments [1-3, 5]

$$I \geq \frac{10^{-3}}{\tau}\left(\frac{1}{(\lambda_v)}\right)^3 \qquad (9).$$

For example, if $\lambda_v$~$10^{-8}$ m (it corresponds to the conditions of the experiment for total pressure about 30bar, calculated supersaturation $S_m$~ 1.53), we have numerical estimation of instantaneous nucleation rate I, which correspond to supersaturation in nucleation zone,

$I \geq 10^{27}$ particles/$m^3$s.

For smaller total pressure, the mean free path of molecule is larger, and therefore, in order to have effective coalescence, according (9) and (2), the nucleation rate may be smaller for observation similar effects. The treatment of experimental data, which is presented in Appendix 1, confirms this conclusion and, therefore, additionally supports out point of view.

As real nucleation rate is very high, will try estimate the value of observed nucleation rate. We can use the calculated value of droplet radius **Re**, droplet with such radius leave the nucleation zone, accordingly the previous chapter. The "screening" effect, described above, works for cylinder with radius **Re**~$3\times10\times10^{-5}$ m. Then (the number droplets Nm crossing the surface unit of the plane below nucleation zone can be written as

$$Nm \sim \frac{1}{\pi Re^2}$$

So at our case, Nm is $3\times10^6$. From the continuity equation for the number density droplets it can be shown, that observed nucleation rate $J_{ob}$ can be calculated as

$$J_{ob} = Nm \times vm / h_n ,$$

where vm is the velocity of droplet crossing the lower boundary of nucleation zone, and $h_n$ is the width of nucleation zone (see chapter Results of simulation). For thermodynamic conditions of [1] and pressure about 40 bars, our calculations show that vm~ 0.001m/s. It follow that observed nucleation rate $J_{ob}$ is about $2\times10^6$ particles/$m^3$s. This value of observed nucleation rate agrees reasonably well with experimental data about critical supersaturation.

If supersaturation will be even higher, and operation stability of DCC permits to realize it experimentally, we can expect oscillatory nucleation [14, 15] at conditions of high-pressure experiments.

## Conclusions

One of the major limitations of the current nucleation experiments in DCC is the coupling between nucleation and droplet growth. Typically, the nuclei have the radii on the order of nanometer, and once formed continue to grow in the conditions of DCC. For detection nucleation events, nucleation experiments rely mainly on light scattering from macroscopic droplets.

The numerical simulation of droplet growth and motion in DCC for thermodynamic condition [1], have shown that if propanol droplet radius reached 32÷35 micron at the middle of chamber height, the observer marked that critical supersaturation exists in DCC. All experimentally observed regularities about behavior of the critical supersaturation versus the total pressure and temperature can be explained from this point of view. The influence on nucleation experiments in DCC of molecular weights of vapor and background gas received molecular- kinetic explanation; see expression (5).

For verification results at high- pressure nucleation experiments, the independent measurements of droplet velocity in DCC can be very useful. The results of such measurements permit to determine of droplet radius versus position and to separate nucleation effects and effects related with droplet growth and motion in DCC.

The low observable nucleation rate versus large supersaturation in nucleation zone is explained by instability of the spatially uniform droplet distribution and the work of coalescence mechanism. The larger transport cross-section of molecule of background gas the smaller total pressure is necessary in order to reach the diffusion regime of droplet growth for radius ~$10^{-8}$m.

For high- pressure nucleation experiments, the diffusion regime of growth of some droplet starts from very small size, in particularly, about ~$10^{-8}$m in [1]. As results of diffusion interaction near growing larger droplet, small droplets disappear. Droplets even with radius about $10^{-5}$m stay in nucleation zone, contrary to low pressure experiments [6]. Therefore relatively large domain of non-uniform of supersaturation arises near growing droplet. Even for moderate total pressure, we expect that diffusion interaction between droplets can play substantial role for explanation of the results of nucleation experiments with high nucleation rate [16].

For observed high nucleation rate (~$10^7$ droplet/$m^3$s), the oscillatory nucleation arises in DCC [14, 15]. But the steady – state nucleation have been observed at experiments [1-3, 5]. Therefore this experimental fact serves as additional confirmation our point of view that the puzzle of high – pressure nucleation in DCC is related to the diffusion interactions near faster growing droplet in nucleation zone.

## Appendix 1

For conditions of experiments [1], the numerical simulation of thermodynamic parameters and droplet growth and motion is presented below. As example, we show only results for data, which have been noted as 363.1K for helium served as background gas in reference [1]. The dependence of dimensionless free energy of cluster formation for saddle point $\Delta\Phi(gc,zc)/kT(zc)$ versus pressure is shown in Fig. A1.

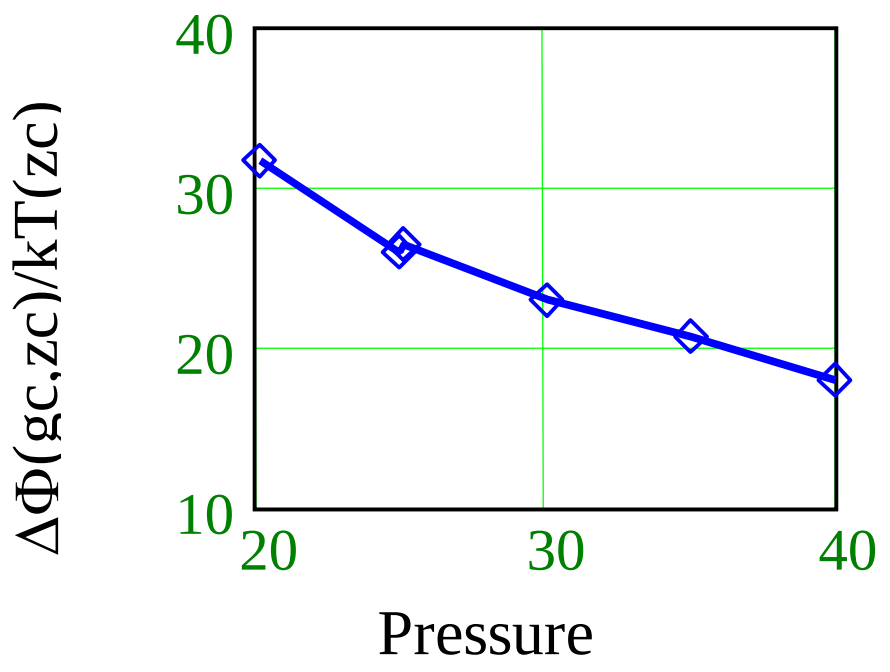

**Figure A1 Dimensionless free energy for the saddle point versus of total pressure (in bar)**

Surprisingly, that the position of the saddle point of dimensionless free energy of cluster formation, where nucleation takes place, changes only 0.562 ÷0.567 of reduced height of chamber. From theoretical point of view, large change of the free energy cluster formation mean that great change of nucleation rate has to be observed. Even for smallest values of the total pressure, $\Delta\Phi(g,z)/kT(z)$ is about 30. Usually, $\Delta\Phi(g,z)/kT(z)$ is about 50 for standard nucleation rate, 1 particles per second per cubic centimeter. For all Figures, the symbols mark the pressure and corresponding temperature of bottom and upper plate [1].

Figure A2 displays the droplet radius at the middle of the chamber versus pressure. For such size of droplet the Knudsen number $Kn_v << 1$. It is remarkable that the radius is practically the same for all experiments. It's mean that the observer see the optical signal the practically the same intensity. The droplet radius is larger that 25 micron, that agree with data from handbook [9]. As discussed above, pressure affects significantly on droplet growth. But during experiments for visualization purposes, the increasing of pressure was practically compensated by increasing the supersaturation near the saddle point [1-3,5]. The constancy of the observed nucleation rate is the effect related with diffusion interaction between small clusters in nucleation zone and "screening" effect near larger growing droplet. Emphasize, the effects of constancy of nucleation rate under increasing of supersaturation cannot be explained in “one-droplet” approximation, used in the vast majority publication devoted to nucleation kinetics in gases.

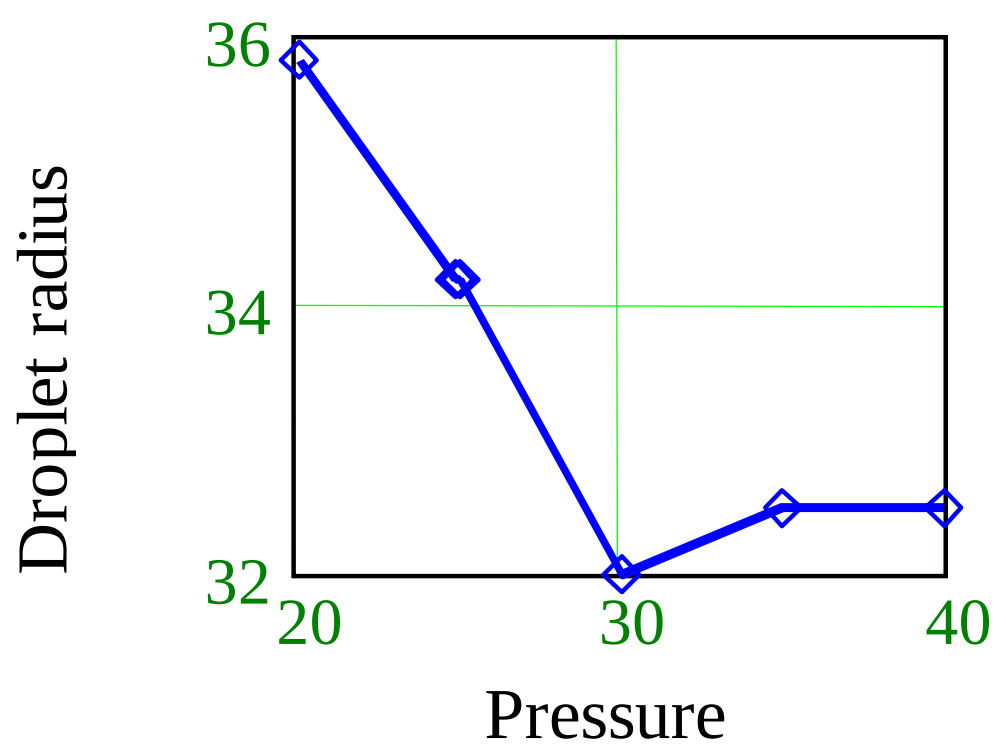

**Figure A2 Calculated droplet (in microns) radius at the middle of chamber versus pressure (in bars).**

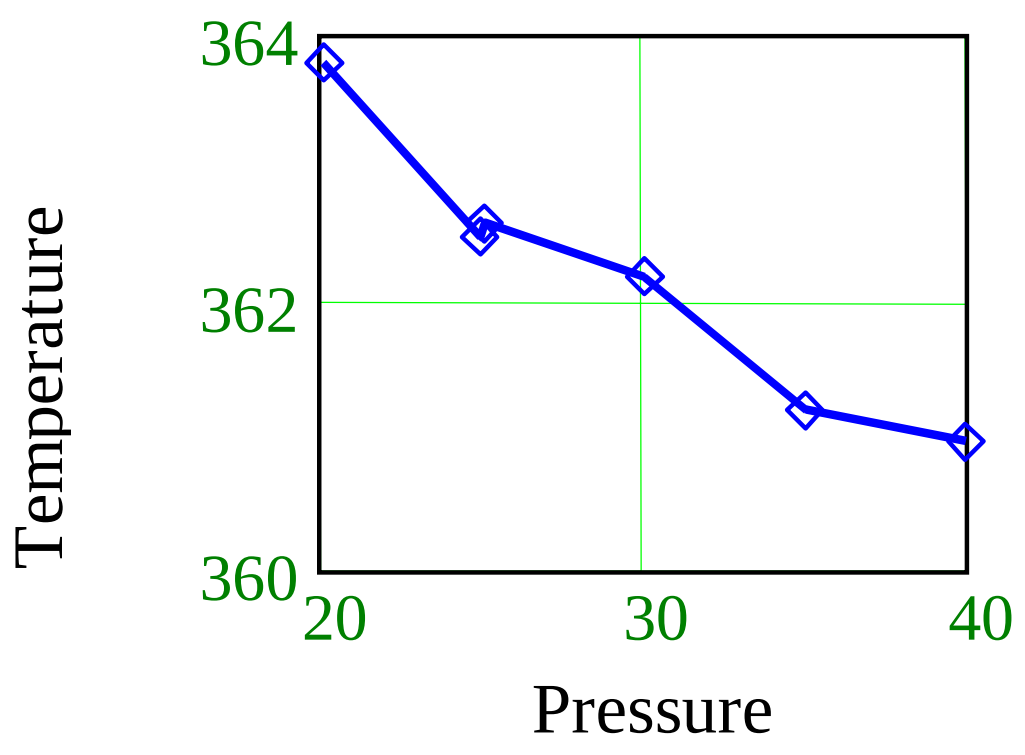

**Figure A3. Temperature at the position of the saddle point versus pressure (in bars)**
for partial set of experimental data (363.1K) from [1]

In Figure A3 the temperature of vapor –helium mixture is shown on the height of the saddle point. For this set of experiments the temperature practically constant with relative accuracy more 1%.

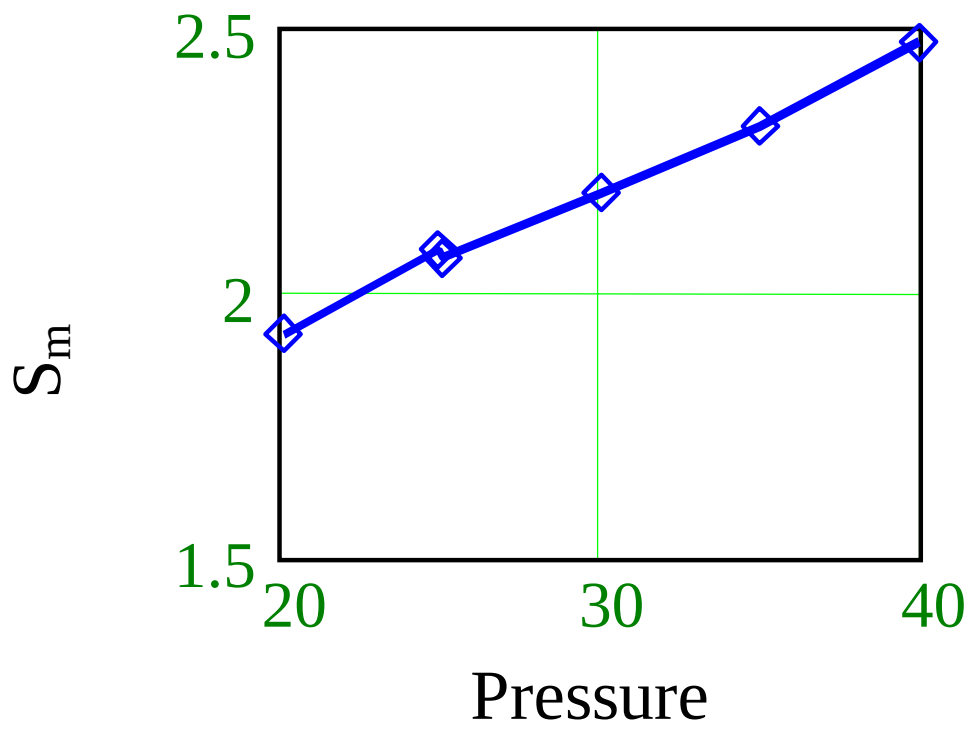

**Figure A4. The maximum of supersaturation Sm versus total pressure P (in bars)**

In Fig.A4 the maximum of supersaturation in DCC versus pressure is presented. To note, that calculated dependence of maximum of supersaturation versus pressure is practically linear function as the experimental dependence in [1].

## Appendix 2

For high- pressure nucleation experiments with background gas pressure much higher than vapor partial pressure, one- dimensional mathematical model for reckoning profiles of vapor density and temperature can be written as the system of two steady-state equations. We assume that the heat and mass transfer processes are essentially one-dimensional, stable, and we neglect the Dufour effect here. The first equation is diffusion equation where diffusion coefficient depends on as total pressure as temperature profile.

$$\partial_z\left[D(T(z))\partial_z n(z)\right]=0 \qquad \text{(A2.1)}$$

The second equation is heat diffusivity equation

$$\partial_z\left[a^2(T(z))\partial_z T(z)\right]=0\,, \qquad \text{(A2.2)}$$

where $a^2$ is the heat diffusivity coefficient, depending on temperature.

The boundary conditions have the standard form for description of the heat and mass transfer processes in DCC:

$$n(0)=n_s(Tb) \qquad \text{(A2.3)}$$

$$n(H)=n_s(Tu) \qquad \text{(A2.4)}$$

and boundary conditions for temperature

$$T(0)=Tb \qquad \text{(A2.5)}$$

$$T(H)=Tu \qquad \text{(A2.6)}$$

Remind that, Tb is the temperature of surface of liquid pool at the bottom of chamber, and Tu is the temperature of liquid film on the surface of upper plate of the DCC.

For DCC more detailed discussion of fine effects related with interaction diffusion, heat conductivity, and hydrodynamic processes are presented at [15].

Steady- state solutions of diffusion and heat diffusion equation with constant coefficients are very good initial approximations to an approximate solutions of equations (A2.1-A2.2) with the same boundary conditions. We seek the approximate solution of (A2.1-A2.2) in the following form by Galerkin's method

$$n(z) = n(0) + \frac{(n(H) - n(0))z}{H} + An_1 \sin\left(\frac{\pi z}{H}\right) \quad \text{(A2.7)}$$

and

$$T(z) = Tb + \frac{(Tu - Tb)z}{H} + At_1 \sin\left(\frac{\pi z}{H}\right)$$

Substituting the expressions of (2.6-2.7) into system of equations ((A2.1-A2.2), multiplying on $\sin(\pi z / H)$ and integrating over all z, we receive the system of algebraic equations for determining unknown coefficient $An_1$ and $At_1$ [10]. For simplicity, contributions of higher modes, related with $\sin(2\pi z / H)$, $\sin(3\pi z / H)$, ets, which are formed the complete set of functions, have been dropped. It can be shown that their contribution is significantly less important than the first mode.

It should be noted, iteration procedure was used for solving of algebraic equations. The linear profile of temperature was at the first iteration for calculation of the integrals of mass transfer coefficients, than new profile of temperature was used for the second iteration, and so forth. For purpose of this paper, we used only two iterations in our calculations of vapor and temperature profiles.

To mention, that the Defour effect can be included into the system of Eqs. (A2.1-A2.2) also, and Galerkin's method of approximate solution can be easily adopted for this advanced version of description of heat and mass transfer processes.